\documentclass[%
 reprint,
 superscriptaddress,
 showpacs,
 showkeys,
 nofootinbib,
  amsmath,amssymb,
 pra,
  longbibliography,
  floatfix,
 ]{revtex4-2}

\usepackage[normalem]{ulem}

\usepackage{adjustbox}
\usepackage{physics}
\usepackage{hyperref}
\usepackage{amsmath}
\usepackage{amssymb}
\usepackage{amsthm}
\usepackage{bm} 
\usepackage{graphicx}

\RequirePackage{times}
\RequirePackage{mathptm}

\usepackage{url}
\usepackage[x11names]{xcolor}
\usepackage{eepic}
\usepackage{tikz}
\usetikzlibrary{decorations.markings}
\usepackage {pgfplots}
\pgfplotsset {compat=1.8}
\usepackage{epstopdf}
\usepackage[normalem]{ulem}
\theoremstyle{definition}

\newcommand{\C}{\mathbb{C}}

\newcommand{\iprod}[2]{\langle #1 | #2 \rangle}

\begin{document}

\title{Binary Quantum Random Number Generator Based on Value Indefinite
Observables}

\author{Cristian S. Calude}
\email{cristian@cs.auckland.ac.nz}
\homepage{http://www.cs.auckland.ac.nz/~cristian}

\affiliation{School of Computer Science, University of Auckland,
Private Bag 92019, Auckland, New Zealand}

\author{Karl Svozil}
\email{karl.svozil@tuwien.ac.at}
\homepage{http://tph.tuwien.ac.at/~svozil}

\affiliation{Institute for Theoretical Physics,
TU Wien,
Wiedner Hauptstrasse 8-10/136,
1040 Vienna,  Austria}

\date{\today}

\begin{abstract}
All quantum random number generators based on measuring value indefinite observables are at least three-dimensional because the Kochen-Specker Theorem and the Located Kochen-Specker Theorem are false in dimension two. In this article, we construct quantum random number generators based on measuring a three-dimensional value indefinite observable that generates binary quantum random outputs with the same randomness qualities as the ternary ones: the outputs are maximally unpredictable.
\end{abstract}


\maketitle

\newpage

\section{Introduction}

In 1946, J.~von Neumann developed a pseudo-random generator (PRNG) with the following algorithm:
``start with an initial random seed value, square it, and slice out the middle digits.''
A sequence obtained by repeatedly using this method exhibits {\it some} statistical properties of randomness.
While the seeds completely determine PRNGs, hundreds of billions of pseudo-random numbers are used daily to encrypt electronic network data.
Their pitfalls have been discovered in the Internet era.
An example is the discovery in 2012 of a weakness in the encryption system RSA~\cite{prng_flow};
the flaw was traced to the numbers a PRNG has produced~\cite{factor_wrong2012}.

New types of random generators have been developed to remedy these flaws, specifically quantum random number generators (QRNGs).
In the last decade, QRNGs proliferated because higher quality randomness is required in many areas, from cryptography,
statistics, and information science to medicine and physics.

QRNGs are considered to be ``better than PRNGs'' because they are based on the
``fundamental unpredictability of well-chosen and controlled quantum processes''~\cite{Quantis2020},
a weak assertion, particularly because it is well-known that the notion of ``true randomness''
interpreted as ``lack of correlations'' or ``maximal randomness'' is mathematically vacuous~\cite{calude:02}.
Can we construct QRNGs ``provably better'' than PRNGs? There are two types of QRNGs ``theoretically certified'':
by  the Bell inequalities~\cite{10.1038/nature09008,NonakaMyriam2022Tros,Hayashi:2022aa}
and by the Located Kochen-Specker Theorem~\cite{2012-incomput-proofsCJ,2015-AnalyticKS},
a form of the Kochen-Specker Theorem, see~\cite{kochen1,Landsman2020,cabello2021contextuality} for detailed reviews.

To date, only the second type of QRNGs has been mathematically proven to be better than \textit{any} PRNG~\cite{2012-incomput-proofsCJ,aguero_trejo_new_2021,RSPA23}.
These QRNGs are three-dimensional:
Since two-dimensional analogs of the Kochen-Specker Theorem as well as the Located Kochen-Specker Theorem are false,
the generated sequences must be at least ternary~\cite{svozil-2009-howto}.
Therefore, to obtain sequences of quantum random bits with the same quality of randomness, we need to apply a ``three-to-two'' symbol conversion algorithm that preserves the level of randomness.
In this article, we pursue an alternative physical conversion:
We construct quantum random generators based on measuring a three-dimensional value indefinite observable,
and operationally---with physical means---generate binary quantum random outputs with the same quality of randomness as the ternary ones.
Such outputs are maximally unpredictable~\cite{2014-nobit}.
Although the results are presented in $\mathbb{C}^3$, they can easily be generalized to $\mathbb{C}^n$ with $n>3$.

\section{Nomenclature and definitions}
\label{2023-viext-defs}

By $n$, we denote a positive integer.
We denote by $\C$ the set of complex numbers and employ the standard quantum mechanical bra-ket notation. In this context, (unit) vectors in the Hilbert space $\C^n$ are represented as $\ket{\cdot}$.
Our focus will be on one-dimensional projection observables.
We denote by $E_\psi$ the operator $E_\psi=\ket{\psi}\bra{\psi}/\vert \iprod{\psi}{\psi}\vert$ projecting the Hilbert space $\C^n$ onto the linear subspace spanned by $\ket{\psi}$.

In the following, we formalize hidden variables and the concept of value definiteness as in~\cite{2015-AnalyticKS}.
Fix $n>1$. Consider $\mathcal{O} \subseteq \{ E_\psi \mid \ket{\psi} \in \mathbb{C}^n \}$, a nonempty set of one-dimensional projection observables on the Hilbert space $\mathbb{C}^n$.
        A set $C \subset \mathcal{O}$ is a {\it context of $\mathcal{O}$} if $C$ has $n$ elements (that is, $|C|=n$), and for all $E_\psi, E_\phi \in C$ with $E_\psi \neq E_\phi$, $\iprod{\psi}{\phi}=0$.

Since distinct one-dimensional projection observables commute if and only if they project onto mutually orthogonal linear subspaces, a context $C$ of $\mathcal{O}$ is a maximal set of compatible one-dimensional projection observables on $\C^n$. Due to the correspondence (up to a phase-shift) between unit vectors and one-dimensional projection observables, a context is uniquely defined by an orthonormal basis of $\C^n$.

A function is partial if it may be undefined for some values; a function
defined everywhere is called total. The square root operation on the real numbers is partial
 because negative real numbers do not have real square roots.
Partial functions were introduced in
computability theory in 1930s~\cite{davis-58}
to model non-halting computations; they were used in quantum physics in~\cite{2012-incomput-proofsCJ}.

        A {\it value assignment function (on $\mathcal{O}$)} is a {\it partial two-valued} function $v: \mathcal{O} \to \{0,1\}$,
assigning values to some (possibly all) observables in $\mathcal{O}$.
While we could allow $v$ to be a function of both the observable $E$ and the context $C$ containing $E$,
enabling contextual value assignments for the sake of compactness,
we define $v$ as a {\it noncontextual} value assignment function; this property is also called {\it contextual independence}.

        An observable $E \in \mathcal{O}$ is {\it value definite (under $v$)} if $v(E)$ is defined;
otherwise, it is {\it value indefinite (under $v$)}.
Similarly, a context $\mathcal{O}$ is value definite (under $v$) if every observable $E \in \mathcal{O}$ is value definite.

Assuming contextual independence, if  $v(E)=1$, the measurement of $E$ in every context containing $E$ must yield the outcome  $1$.
More generally, every value (in)definite observable $E$ in one context must also value (in)definite in all other contexts containing $E$.
This unique value, $0$, $1$, or undefined, depends on a particular state preparation and a specific collection of observables and contexts, which can be compactly represented by a hypergraph~\cite{greechie:71,Bretto-MR3077516} (for more details, see later Section~\ref{2023-viext-qrngdefs}).

Let $\mathcal{O}$ be a set of one-dimensional projection observables on $\C^n$, and let $v:\mathcal{O}\to\{0,1\}$ be a value assignment function.
Then, $v$ is {\it admissible} if the following two conditions hold for every context $C$ of $\mathcal{O}$:
        \begin{itemize}
                \item[(a)] Exclusivity: If there exists an $E\in C$ with $v(E)=1$, then $v(E')=0$, for all $E'\in C\setminus\{E\}$.
                \item[(b)] Completeness:                  If there exists an $E\in C$ with $v(E')=0$,   for all $E'\in C\setminus\{E\}$, then $v(E)=1$.
        \end{itemize}

Admissibility is a weaker requirement than the usual assumption of the existence
of a two-valued state---a total value assignment---because fewer than $n-1$ elements in a context on $\C^n$ may be assigned the value 0, and no element is assigned the value 1. If the value assignment is partial, then the observables corresponding to these remaining elements are value indefinite.

For example, in $\C^3$, consider a  context that has no element with either value $0$ or $1$ (and thus the value assignments of all three elements are undefined) and another context that has only a single element that is assigned the value $0$, and the other two undefined.

However, if the value assignment on a particular set $\mathcal{O}$ of one-dimensional projection observables on $\C^n$ is total,
then admissibility coincides with the standard definition of two-valued state(s).

Admissibility permits undefined values, and thus value indefiniteness of an observable $E$ if both outcomes (0 and 1) of a measurement of $E$
are incompatible with the definite values of other observables sharing a context with $E$.
An explicit construction of such a configuration has been presented in Tables~I and~II and Figure~2 in~\cite{2015-AnalyticKS}, also appearing in Figure~\ref{2023-viext-TIFS&TITSconcatenated}.

If $v(E)=1$, the measurement of every observable in
every
context $C$ containing $E$ must yield the outcome 1 for $E$.
Consequently, to avoid contradiction, the outcomes of measurements for all the other observables in the context must be 0, and vice versa.  On the other hand, if $v(E)=0$,
then the measurements of the other observables in $C$ could yield the values 1 and 0 (as long as only one yields 1).

\section{Three-Dimensional QRNGs}

This section introduces the physical principles and assumptions on which the notion of being ``better than any PRNG'' operates~\cite{2012-incomput-proofsCJ,aguero_trejo_new_2021,RSPA23}.
We then proceed to an explicit example based on a configuration of observables that realizes a QRNG according to these principles.

\subsection{Principles of Three-Dimensional QRNGs}

In the articles~\cite{2008-cal-svo,2012-incomput-proofsCJ,2015-AnalyticKS}, the following protocol was used to construct a class of  3-dimensional QRNGs:
\begin{quote} {\it repeatedly locate a value indefinite observable in $C^3$, measure it and record the output}.
\end{quote}
 The Kochen-Specker Theorem~\cite{kochen1} guarantees only the existence of value indefinite
observables, so the above protocol cannot use it. In contrast, the located version of the theorem~\cite{2008-cal-svo,2012-incomput-proofsCJ} allows the construction of value indefinite observables, which can then be measured. In detail, consider a quantum system
 described by the state
$\ket{\psi}$ in a Hilbert space ${\mathbf C}^n$, $n\ge 3$ and choose  a value indefinite observable (quantum state) $\ket{\phi}$ that is neither orthogonal nor parallel to $\ket{\psi}$ {\rm ($0<\abs{\bra{\psi}\ket{\phi}}<1$)}. If the following three conditions are satisfied:

\begin{enumerate}
\item  admissibility, as defined in Section~\ref{2023-viext-defs}, \item non-contextuality,
the outcome obtained by
measuring a value definite observable
  does not
depend on other compatible
 observables which may be measured alongside it, and \item
Eigenstate principle, if a quantum system is prepared in the
state $\ket{\psi}$,
then the projection observable
$P_\psi$ is value definite,
\end{enumerate}
then the projection observable $P_\phi$ is value indefinite.

Furthermore, in~\cite{RSPA23}, it was proved that
 given every probability distribution $(p_1,p_2,p_3)$ ($\sum_{i}p_i=1$ and $0\le  p_i < 1$), a value indefinite quantum state can be constructed which, by a {\it universal} measurement, produces the
outcomes with probabilities $p_i$.

 The {\it universal} measurement is described
by the unitary operator given by the unitary matrix~\cite{RSPA23}:\footnote{This is obtained in terms of the spin-1 operator in the $x$-direction $S_x$, and its associated unit eigenvectors (through its spectral decomposition).}

\begin{equation}
U_x =\frac{1}{2}
\begin{pmatrix}
1 & \sqrt{2} & 1\\
\sqrt{2} & 0 & -\sqrt{2}\\
1 & -\sqrt{2} & 1
\end{pmatrix}.
\label{2023-ux}
\end{equation}

\noindent The quantum state
(modular phase factors)

\begin{equation}
\label{vi}
\vert \psi \rangle=\begin{pmatrix}\sqrt{p_1},\sqrt{p_2},\sqrt{p_3}\end{pmatrix}^T,
\end{equation}
is value indefinite~\cite[Theorem 4.1.]{RSPA23} and the result of the measurement of $\vert \psi \rangle$ on $U_x$ with respect to the Cartesian standard basis produces the outcome
 $i\in \{0,1,2\}$ with probability $p_i$.

In fact, {\it every unitary operator is universal with respect to the value indefinite quantum state} $\vert \psi \rangle$. This is easy to see for
the identity matrix, the most elementary case. 
As every arbitrary unitary operator $U$  can be written in terms of two orthonormal bases,
$\{ \vert f_1 \rangle, \vert f_2 \rangle , \vert f_3 \rangle  \}$
and the  Cartesian standard basis $\{ \vert e_1 \rangle, \vert e_2 \rangle , \vert e_3 \rangle  \}$,
as
$$U=
\vert f_1\rangle \langle e_1\vert +
\vert f_2\rangle \langle e_2\vert +
\vert f_3\rangle \langle e_3\vert ,$$
 we have by {\ref{vi})
\begin{equation}
\vert \psi \rangle =
\sqrt{p_1}\vert e_1 \rangle + \sqrt{p_2}\vert e_2 \rangle + \sqrt{p_3}\vert e_3 \rangle.
\end{equation}
If we measure the value indefinite $\vert \psi \rangle$ by $U$
in terms of the orthonormal basis
$\{ \vert f_1 \rangle, \vert f_2 \rangle , \vert f_3 \rangle  \}$
}
we get the outcome
 $i\in \{0,1,2\}$ with probability $p_i$.
 If we  measure the output of $U$ in terms of the
Cartesian standard basis $\{ \vert e_1 \rangle, \vert e_2 \rangle , \vert e_3 \rangle  \}$, then the input state has to be pre-processed: 
$U^{-1}\vert \psi \rangle = U^\dagger \vert \psi \rangle$, where $\dagger$ stands for the  Hermitian adjoint.

Finally, using the main result in~\cite{acs-2015-info6040773}, running the above quantum protocol indefinitely, we {\it always} obtain a maximally unpredictable ternary sequence.

The first 3-dimensional QRNG~\cite{2012-incomput-proofsCJ} was constructed by (a)
choosing the quantum state $\vert a \rangle=\begin{pmatrix}0,1,0
\end{pmatrix}^T$---which is value definite with respect to any context containing the observable $\vert a \rangle \langle a\vert = \text{diag}\begin{pmatrix}1,0,0
\end{pmatrix}$ because $\vert a \rangle$ is not in the context formed by the row vectors of $U_{x}$, (b)  choosing a quantum state that is neither orthogonal nor parallel to
it and (c)  applying the measurement (\ref{2023-ux}).  The probabilities of the outputs 0, 1, and 2 generated by this quantum random generator are $\frac{1}{2},0  $  and $\frac{1}{2}$,  respectively, so theoretically, every sequence generated by this protocol is binary.

 Does the probability 0 output
 endanger the applicability of the Kochen-Specker Theorem (see also the
 principle of three and higher-dimensionality of QRNGs~\cite{svozil-2009-howto})? The experimental analysis~\cite{Abbott_2019}, based on the experiments reported in~\cite{Kulikov2017Dec}, suggested that the answer to the question posed in~\cite{Arkady_Fedorov-pc},  is negative.\footnote{A very small number of outputs 2 have been obtained.} We can now provide a theoretical negative answer using the {\it universal} measurement (\ref{2023-ux}) to value indefinite quantum states.

By  changing the input quantum state $\vert a \rangle = \begin{pmatrix} 0, 1 , 0  \end{pmatrix}^T$
  to $\vert a \rangle = \begin{pmatrix} 1 , 0 , 0 \end{pmatrix}^T$ and using the  measurement (\ref{2023-ux})
 we obtain ternary quantum random numbers 0,1,2 generated with with probabilities $1/2,1/4,1/4$, respectively,  hence
  ``genuine'' ternary sequences.

 As many current applications require random binary sequences, in~\cite{RSPA23},
 the computable  alphabetic morphism $\varphi \colon \{0,1,2\} \rightarrow \{0,1\}$
 \begin{equation}
 \varphi(x)=
 \begin{cases}1,&\text{if }x=0,
 \\0,&\text{if }x=1,
 \\0, &\text{if } x=2,\end{cases}
\label{2023-alphabeticmorphism}
 \end{equation}

  A slightly modified version of this alphabetic morphism
 was used to
 transform ternary sequences into binary ones and preserve their maximal unpredictability
 for the probability distributions  $\frac{1}{4},\frac{1}{2},\frac{1}{4}$ and $\frac{1}{2},\frac{1}{2}$, respectively;
see~\cite{CALUDE202131} and Section~7 in~\cite{aguero_trejo_new_2021}. Can we ``quantize'' the algorithmic post-processing (\ref{2023-alphabeticmorphism})?

Quantum mechanically, this alphabetic morphism corresponds to a post-processing of the output of $U_x\vert a \rangle$.
In general, by
post-processing of a  unitary transformation $A$ we mean the unitary transformation $B = U'A$, where $U'$ is a suitable unitary transformation.
Physically, this corresponds to the serial composition of beam splitters, first applying $A$ and then $U'$.

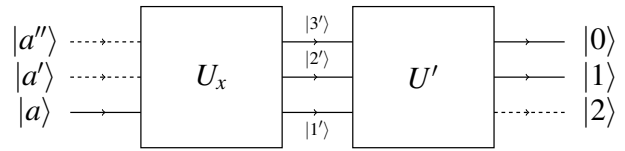
\begin{figure}
\begin{center}
\resizebox{.46\textwidth}{!}{
\begin{tikzpicture}  [scale=1]
\tikzstyle{every path}=[line width=1pt]

\tikzset{->-/.style={decoration={
  markings,
  mark=at position #1 with {\arrow{>}}},postaction={decorate}}}

\draw (0,0) rectangle (4,4);
\draw (6,0) rectangle (10,4);
\draw[->-=.5] (4,3) -- (6,3);
\draw[->-=.5] (4,2) -- (6,2);
\draw[->-=.5] (4,1) -- (6,1);
\draw[->-=.5,dashed] (-2,3) -- (0,3);
\draw[->-=.5,dashed] (-2,2) -- (0,2);
\draw[->-=.5] (-2,1) -- (0,1);
\draw[->-=.5] (10,3) -- (12,3);
\draw[->-=.5] (10,2) -- (12,2);
\draw[->-=.5,dashed] (10,1) -- (12,1);
\node at (2,2) {\Huge $U_x$};
\node at (8,2) {\Huge $U'$};

\node at (-3,3) {\Huge $\vert a'' \rangle$};
\node at (-3,2) {\Huge $\vert a' \rangle$};
\node at (-3,1) {\Huge $\vert a \rangle$};
\node at (5,3.5) {\Large $\vert 3' \rangle$};
\node at (5,2.5) {\Large $\vert 2' \rangle$};
\node at (5,0.5) {\Large $\vert 1' \rangle$};
\node at (13,3) {\Huge $\vert 0 \rangle$};
\node at (13,2) {\Huge $\vert 1 \rangle$};
\node at (13,1) {\Huge $\vert 2 \rangle$};
\end{tikzpicture}
}
\end{center}
                \caption{A horizontal schema of two beam splitters $U_x$ and $U'$ in serial composition $U'U_x$,
with the `input' state prepared in $\vert a\rangle$,
and two `active output' ports in states $\vert 0\rangle$ and $\vert 1\rangle$.}
\label{2023-viext-souubone2}
\end{figure}

The post-processing of (\ref{2023-alphabeticmorphism}) results in the `merging'
of a state with three nonzero components
(or coordinates with respect to a particular basis, here the Cartesian standard basis)
into a state with two
nonzero components. The merging is justified only if the corresponding input ports belong to the same context.
In other words, the corresponding observables have mutually exclusive outcomes---a condition satisfied by a beam splitter realizing $U_x$.
The schema is presented in Figure~\ref{2023-viext-souubone2}.
Thereby,  the unitary matrix is

\begin{equation}
U' =\frac{1}{2\sqrt2}
\begin{pmatrix}
1+\sqrt2& \sqrt2 & 1-\sqrt2\\
1-\sqrt2 & \sqrt2 & 1+\sqrt2 \\
\sqrt2 & -2 & \sqrt2
\end{pmatrix}
\label{2023-viext-upux}
\end{equation}

\noindent corresponds to the alphabetic morphism $\varphi$.
Then, the combined transformation is
\begin{equation}
U'U_x =
\frac{1}{\sqrt{2}}
\begin{pmatrix}
1&1&0\\
1&-1&0\\
0&0&\sqrt{2}
\end{pmatrix}
.
\label{2023-viext-u}
\end{equation}

This unitary matrix $U'U_x$ corresponds to a beam splitter configuration that first allows a state $\vert a\rangle$ to be `expanded' by a unitary matrix $U_x$ with three nonzero components. Simultaneously,
given $\vert a\rangle$, the output state $U_x\vert a\rangle$ is a value-indefinite observable `merged' or `folded' by the unitary matrix $U'$,
representing a serially concatenated beam splitter that transforms this state
into one with two nonzero components of equal probability amplitudes.
On input $\vert a\rangle $ the unitary transformation  $U'U_x$ generates a ternary output
with the probability distribution
$\begin{pmatrix}\frac{1}{2},\frac{1}{2},0\end{pmatrix}$, which corresponds to the binary output with the probability distribution
$\begin{pmatrix}\frac{1}{2},\frac{1}{2}\end{pmatrix}$.

How can we realize this transformation in terms of unitary equivalence?
Two transformations, $A$ and $B$, are unitarily equivalent if there exists a unitary matrix $V$ such that $B = V^\dagger AV$, where
$V^\dagger$ means the Hermitian adjoint, or conjugate transpose, of $V$.
If $V$ is real-valued then $V^\dagger =V^T$ is just the transpose $V^T$ of $V$.

From Specht's Theorem~\cite{Horn-Johnson-MatrixAnalysis,Dokovi__2007}, two unitary matrices are unitary equivalent if their eigenvalues coincide.
In our case, both $U_x$ in~(\ref{2023-ux}) as well as $U'U_x$ in~(\ref{2023-viext-upux}) have one eigenvalue $-1$, and a double eigenvalue $1$.
More explicitly, the matrix

\begin{equation}
V=
\begin{pmatrix}
 \frac{1}{2\sqrt{3}} \sqrt{2-\sqrt{2+\sqrt{3}}}
   & \frac{1}{2\sqrt{3}} \sqrt{2+\sqrt{2+\sqrt{3}}} & \sqrt{\frac{2}{3}} \\
 -\frac{1}{\sqrt{6}}\sqrt{2-\sqrt{2+\sqrt{3}}} &
   -\frac{1}{\sqrt{6}} \sqrt{2+\sqrt{2+\sqrt{3}}} &
   \frac{1}{\sqrt{3}} \\
 \frac{1}{2} \sqrt{2+\sqrt{2+\sqrt{3}}} & -\frac{1}{2}
   \sqrt{2-\sqrt{2+\sqrt{3}}} & 0
\end{pmatrix}
\end{equation}
satisfies the equality  $V^TU_xV=U'U_x$: this proves
 that the matrix $U_x$ defined in (\ref{2023-ux}) is unitarily equivalent to the matrix combination $U'U_x$ in (\ref{2023-viext-u}).

\subsection{Configuration of Observables Realizing the Principles of Three-Dimensional QRNGs}

For the sake of an example, take a configuration of observables enumerated  in~\cite[Table~I]{2018-minimalYIYS}
presented in Figure~\ref{2023-viext-TIFS&TITSconcatenated},  as $v(a)=1$,
in the context $\{b,2,3\}$, the observable $2$ is value definite with $v(2)=0$, whereas
both observables
$b$ and $3$ are value indefinite.
Therefore, not all elements of $C\setminus \{E\}$ need to be value indefinite:
Indeed, in the context $\{b,2,3\}$, the observable $b$ is value indefinite.
But from the two remaining elements in $\{b,2,3\} \setminus \{b\}=\{2,3\}$,
$2$ is value definite with $v(2)=0$,
and $3$ is value indefinite.

\medskip

For the sake of an example, we shall use a hypergraph introduced in~\cite{2015-AnalyticKS}
and split it into segments serving as true-implies-false (TIFS) and true-implies-true (TITS) gadgets~\cite{2018-minimalYIYS}.

\begin{figure*}
                \begin{center}
\begin{tabular}{ccc}
\resizebox{.46\textwidth}{!}{
                \begin{tikzpicture}  [scale=0.5]

                \tikzstyle{every path}=[line width=2pt]
                \tikzstyle{c1}=[draw,black,fill=green,circle,inner sep={3},minimum size=6]
                \tikzstyle{s1}=[draw,black,fill=red,rectangle,inner sep={4},minimum size=7]
                \tikzstyle{l1}=[draw=none,circle,minimum size=4]


                \draw (4,0) coordinate[c1,label=180:{\Large $b$}] (b)
                        -- (13,0) coordinate[c1,label=45:$2$] (2)
                        -- (22,0) coordinate[s1,label=0:$3$] (3)
                        -- (26,12) coordinate[c1,pos=0.8,label=0:${21}$] (21) coordinate[c1,label=0:${23}$] (23);
                        \draw  (22,18.5) coordinate[c1,label=0:$5$] (5)
                        -- (13,18.5) coordinate[s1,label=90:{\Large $a$}] (a)
                        -- (4,18.5) coordinate[c1,label=90:$4$] (4)
                        -- (0,12) coordinate[c1,pos=0.6,label=180:${10}$] (10) coordinate[s1,label=180:$7$] (7)
                        -- (b) coordinate[c1,pos=0.2,label=180:$6$] (6);

                \draw (a) -- (2) coordinate[c1,pos=0.5,label=315:$1$] (1);

                \draw (5) -- (22,6) coordinate[s1,pos=0.4,label=0:${11}$] (11) coordinate[c1,label=0:$9$] (9)
                        -- (b) coordinate[s1,pos=0.3,label=280:$8$] (8);

                \draw (4) -- (4,6) coordinate[s1,pos=0.4,label=180:${28}$] (28) coordinate[c1,label=180:${22}$] (22)
                        -- (3) coordinate[c1,pos=0.3,label=260:${19}$] (19);

                \coordinate (25) at ([xshift=-4cm]1);
                \coordinate (27) at ([xshift=4cm]1);

                \draw (22) -- (25) coordinate[c1,pos=0.5,label=115:${24}$] (24) coordinate[s1,label=270:${25}$] (25)
                        -- (9) coordinate[c1,pos=0.8,label=90:${35}$] (35);

                \draw (7) -- (27) coordinate[c1,pos=0.5,label=90:${34}$] (34) coordinate[c1,label=90:${27}$] (27)
                        -- (23) coordinate[s1,pos=0.3,label=270:${26}$] (26);

                \draw (10) -- (15.5,17.5) coordinate[c1,pos=0.5,label=90:${12}$] (12) coordinate[s1,label=15:${13}$] (13);

                \draw (28) -- (10.5,15) coordinate[c1,pos=0.5,label=90:${30}$] (30) coordinate[c1,label=90:${15}$] (15)
                        -- (11) coordinate[c1,pos=0.6,label=90:${14}$] (14);

                \draw (15) -- (1) coordinate[s1,pos=0.2,label=15:${17}$] (17)
                        -- (13) coordinate[c1,pos=0.3,label=0:${16}$] (16);

                \draw (19) -- (16) coordinate[s1,pos=0.3,label=180:${18}$] (18)
                        -- (8) coordinate[c1,pos=0.7,label=180:${32}$] (32);

                \draw (6) -- (17) coordinate[c1,pos=0.7,label=90:${33}$] (33)
                        -- (21) coordinate[c1,pos=0.4,label=90:${20}$] (20);

                \draw (25) -- (1) -- (27);

                \coordinate (ContextLabel) at ([shift=({-2cm,-3mm})]1);

                \end{tikzpicture}
                }
&$\qquad$&
\resizebox{.46\textwidth}{!}{
                \begin{tikzpicture}  [scale=0.5]

                \tikzstyle{every path}=[line width=2pt]
                \tikzstyle{c1}=[draw,black,fill=green,circle,inner sep={3},minimum size=6]
                \tikzstyle{s1}=[draw,black,fill=red,rectangle,inner sep={4},minimum size=7]
                \tikzstyle{l1}=[draw=none,circle,minimum size=4]


                \draw (4,0) coordinate[s1,label=180:{\Large $b$}] (b)
                        -- (13,0) coordinate[c1,label=45:$2$] (2)
                        -- (22,0) coordinate[c1,label=0:$3$] (3);
\draw [dashed] (3) -- (26,12) coordinate[c1,solid,pos=0.8,label=0:${21}$] (21) coordinate[c1,solid,label=0:${23}$] (23);
                        \draw  (22,18.5) coordinate[c1,label=0:$5$] (5)
                        -- (13,18.5) coordinate[s1,label=90:{\Large $a$}] (a)
                        -- (4,18.5) coordinate[c1,label=90:$4$] (4)
                        -- (0,12) coordinate[s1,pos=0.6,label=180:${10}$] (10) coordinate[c1,label=180:$7$] (7)
                        -- (b) coordinate[c1,pos=0.2,label=180:$6$] (6);

                \draw (a) -- (2) coordinate[c1,pos=0.5,label=315:$1$] (1);

                \draw (5) -- (22,6) coordinate[s1,pos=0.4,label=0:${11}$] (11) coordinate[c1,label=0:$9$] (9)
                        -- (b) coordinate[c1,pos=0.3,label=280:$8$] (8);

                \draw (4) -- (4,6) coordinate[c1,pos=0.4,label=180:${28}$] (28) coordinate[s1,label=180:${22}$] (22)
                        -- (3) coordinate[c1,pos=0.3,label=260:${19}$] (19);

                \coordinate (25) at ([xshift=-4cm]1);
                \coordinate (27) at ([xshift=4cm]1);

                \draw (22) -- (25) coordinate[c1,pos=0.5,label=115:${24}$] (24) coordinate[c1,label=270:${25}$] (25)
                        -- (9) coordinate[s1,pos=0.8,label=90:${35}$] (35);

                \draw (7) -- (27) coordinate[c1,pos=0.5,label=90:${34}$] (34) coordinate[s1,label=90:${27}$] (27)
                        -- (23) coordinate[c1,pos=0.3,label=270:${26}$] (26);

                \draw (10) -- (15.5,17.5) coordinate[c1,pos=0.5,label=90:${12}$] (12) coordinate[c1,label=15:${13}$] (13);

                \draw (28) -- (10.5,15) coordinate[s1,pos=0.5,label=90:${30}$] (30) coordinate[c1,label=90:${15}$] (15)
                        -- (11) coordinate[c1,pos=0.6,label=90:${14}$] (14);

                \draw (15) -- (1) coordinate[s1,pos=0.2,label=15:${17}$] (17)
                        -- (13) coordinate[s1,pos=0.3,label=0:${16}$] (16);

                \draw (19) -- (16) coordinate[c1,pos=0.3,label=180:${18}$] (18)
                        -- (8) coordinate[c1,pos=0.7,label=180:${32}$] (32);

                \draw (6) -- (17) coordinate[c1,pos=0.7,label=90:${33}$] (33)
                        -- (21) coordinate[c1,pos=0.4,label=90:${20}$] (20);

                \draw (25) -- (1) -- (27);

                \coordinate (ContextLabel) at ([shift=({-2cm,-3mm})]1);

                \end{tikzpicture}
                }
\\
\\
(a)&&(b)
\end{tabular}
                \end{center}
                \caption{The TIFS gadget hypergraph for $b$ given $v(a)=1$, as well as the TITS gadget hypergraph for $3$ given $v(a)=1$,
illustrates the orthogonality relations among vector labels of the elements of hyperedges~\cite{lovasz-79} within a subset of quantum observables---also known as a faithful orthogonal representation~\cite{Portillo-2015} or coordinatization~\cite{Pavii2018}, as enumerated in~\cite[Table~I]{2018-minimalYIYS}. Red squares represent the value 1, and green circles represent the value 0.
(a) A singular, consistent value assignment is obtained by assuming $v(a)=1$ and $v(b)=0$ and applying the admissibility rules successively~\cite[Figure~(24.2.a)]{Svozil-2018-p}.
(b) An inconsistent value assignment is obtained by assuming $v(a)=v(b)=1$ and applying the admissibility rules successively:
 the context $\{3,21,23\}$, shown dotted, contains three observables with the value 0; hence no admissible value assignment $v$
with the constraint on the input and output ports $v(a)=v(b)=1$ exists. Therefore, if $a$ has a preselected input state $v(a)=1$,
then the value assignment $v(b)$ for the output state $b$ has either to be 0 or needs to be undefined, that is, $b$ is value indefinite.
}
                \label{2023-viext-TIFS}
        \end{figure*}
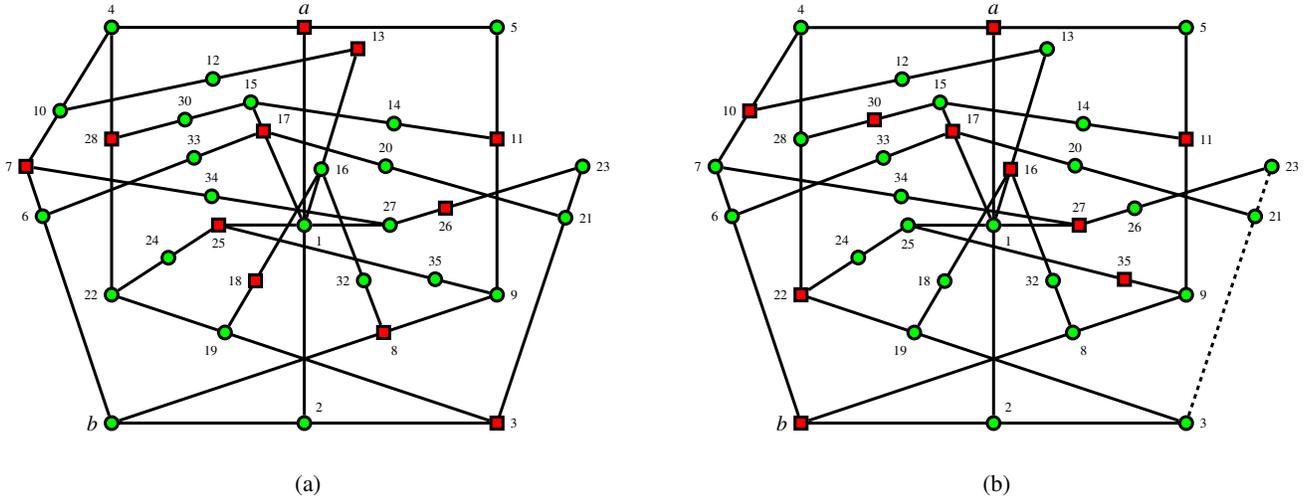

The hypergraph corresponding to the TIFS gadget  in Figure~\ref{2023-viext-TIFS}
illustrates the orthogonality relations among vector labels of the elements of hyperedges~\cite{lovasz-79}, as detailed in~\cite[Table~I]{2018-minimalYIYS}.
By subsequently applying the admissibility rules~\cite[Figure~(24.2.a)]{Svozil-2018-p} a single consistent value assignment,
as in Figure~(\ref{2023-viext-TIFS}a) allows $v(a)=1$ and $v(b)=0$, whereas
an inconsistent value assignment arises when assuming $v(a)=v(b)=1$.
Therefore, for any such configuration of quantum observables, there exists no classical admissible value assignment $v$
satisfying the constraint on the input and output ports $v(a)=v(b)=1$.
Consequently, if $a$ has a preselected input state $v(a)=1$, then
the value assignment $v(b)$ for the output state $b$ cannot be 1.
Therefore, $v(b)$ can only be 0 or undefined.
In the latter case, $b$
is value indefinite.

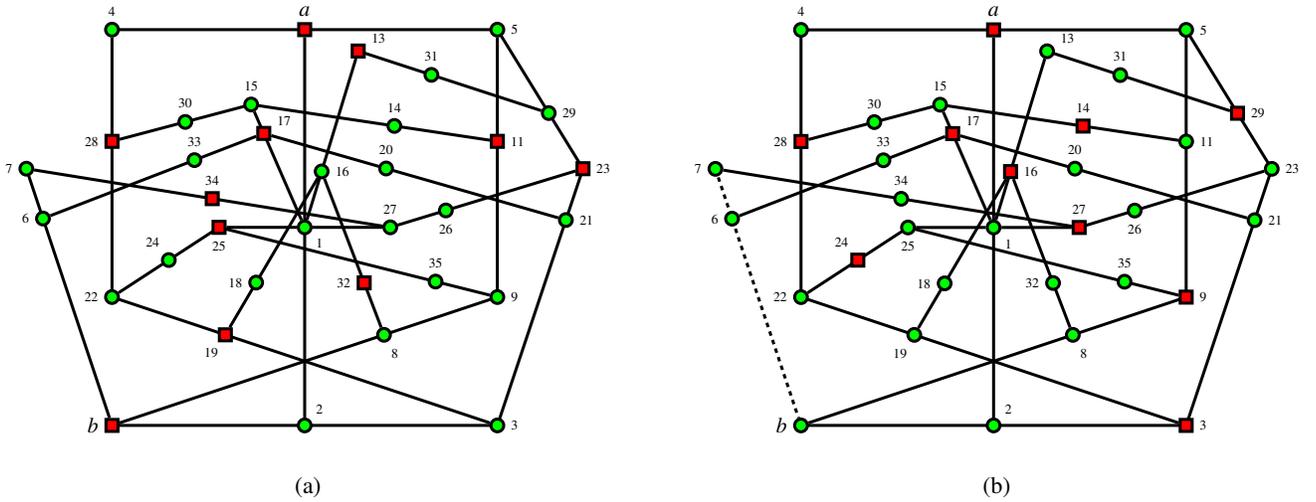
\begin{figure*}
                \begin{center}
\begin{tabular}{ccc}
\resizebox{.46\textwidth}{!}{
                \begin{tikzpicture}  [scale=0.5]

                \tikzstyle{every path}=[line width=2pt]
                \tikzstyle{c1}=[draw,black,fill=green,circle,inner sep={3},minimum size=6]
                \tikzstyle{s1}=[draw,black,fill=red,rectangle,inner sep={4},minimum size=7]
                \tikzstyle{l1}=[draw=none,circle,minimum size=4]


                \draw (4,0) coordinate[s1,label=180:{\Large $b$}] (b)
                        -- (13,0) coordinate[c1,label=45:$2$] (2)
                        -- (22,0) coordinate[c1,label=0:$3$] (3)
                        -- (26,12) coordinate[c1,pos=0.8,label=0:${21}$] (21) coordinate[s1,label=0:${23}$] (23)
                        -- (22,18.5) coordinate[c1,pos=0.4,label=0:${29}$] (29) coordinate[c1,label=0:$5$] (5)
                        -- (13,18.5) coordinate[s1,label=90:{\Large $a$}] (a)
                        -- (4,18.5) coordinate[c1,label=90:$4$] (4);
\draw (0,12)  coordinate[c1,label=180:$7$] (7)
                        -- (b) coordinate[c1,pos=0.2,label=180:$6$] (6);

                \draw (a) -- (2) coordinate[c1,pos=0.5,label=315:$1$] (1);

                \draw (5) -- (22,6) coordinate[s1,pos=0.4,label=0:${11}$] (11) coordinate[c1,label=0:$9$] (9)
                        -- (b) coordinate[c1,pos=0.3,label=280:$8$] (8);

                \draw (4) -- (4,6) coordinate[s1,pos=0.4,label=180:${28}$] (28) coordinate[c1,label=180:${22}$] (22)
                        -- (3) coordinate[s1,pos=0.3,label=260:${19}$] (19);

                \coordinate (25) at ([xshift=-4cm]1);
                \coordinate (27) at ([xshift=4cm]1);

                \draw (22) -- (25) coordinate[c1,pos=0.5,label=115:${24}$] (24) coordinate[s1,label=270:${25}$] (25)
                        -- (9) coordinate[c1,pos=0.8,label=90:${35}$] (35);

                \draw (7) -- (27) coordinate[s1,pos=0.5,label=90:${34}$] (34) coordinate[c1,label=90:${27}$] (27)
                        -- (23) coordinate[c1,pos=0.3,label=270:${26}$] (26);

                \draw  (15.5,17.5) coordinate[s1,label=15:${13}$] (13)
                        -- (29) coordinate[c1,pos=0.4,label=90:${31}$] (31);

                \draw (28) -- (10.5,15) coordinate[c1,pos=0.5,label=90:${30}$] (30) coordinate[c1,label=90:${15}$] (15)
                        -- (11) coordinate[c1,pos=0.6,label=90:${14}$] (14);

                \draw (15) -- (1) coordinate[s1,pos=0.2,label=15:${17}$] (17)
                        -- (13) coordinate[c1,pos=0.3,label=0:${16}$] (16);

                \draw (19) -- (16) coordinate[c1,pos=0.3,label=180:${18}$] (18)
                        -- (8) coordinate[s1,pos=0.7,label=180:${32}$] (32);

                \draw (6) -- (17) coordinate[c1,pos=0.7,label=90:${33}$] (33)
                        -- (21) coordinate[c1,pos=0.4,label=90:${20}$] (20);

                \draw (25) -- (1) -- (27);

                \coordinate (ContextLabel) at ([shift=({-2cm,-3mm})]1);

                \end{tikzpicture}
                }
&$\qquad$&
\resizebox{.46\textwidth}{!}{
                \begin{tikzpicture}  [scale=0.5]

                \tikzstyle{every path}=[line width=2pt]
                \tikzstyle{c1}=[draw,black,fill=green,circle,inner sep={3},minimum size=6]
                \tikzstyle{s1}=[draw,black,fill=red,rectangle,inner sep={4},minimum size=7]
                \tikzstyle{l1}=[draw=none,circle,minimum size=4]


                \draw (4,0) coordinate[c1,label=180:{\Large $b$}] (b)
                        -- (13,0) coordinate[c1,label=45:$2$] (2)
                        -- (22,0) coordinate[s1,label=0:$3$] (3)
                        -- (26,12) coordinate[c1,pos=0.8,label=0:${21}$] (21) coordinate[c1,label=0:${23}$] (23)
                        -- (22,18.5) coordinate[s1,pos=0.4,label=0:${29}$] (29) coordinate[c1,label=0:$5$] (5)
                        -- (13,18.5) coordinate[s1,label=90:{\Large $a$}] (a)
                        -- (4,18.5) coordinate[c1,label=90:$4$] (4);
\draw [dashed](0,12)  coordinate[c1,solid,label=180:$7$] (7)
                        -- (b) coordinate[c1,solid,pos=0.2,label=180:$6$] (6);

                \draw (a) -- (2) coordinate[c1,pos=0.5,label=315:$1$] (1);

                \draw (5) -- (22,6) coordinate[c1,pos=0.4,label=0:${11}$] (11) coordinate[s1,label=0:$9$] (9)
                        -- (b) coordinate[c1,pos=0.3,label=280:$8$] (8);

                \draw (4) -- (4,6) coordinate[s1,pos=0.4,label=180:${28}$] (28) coordinate[c1,label=180:${22}$] (22)
                        -- (3) coordinate[c1,pos=0.3,label=260:${19}$] (19);

                \coordinate (25) at ([xshift=-4cm]1);
                \coordinate (27) at ([xshift=4cm]1);

                \draw (22) -- (25) coordinate[s1,pos=0.5,label=115:${24}$] (24) coordinate[c1,label=270:${25}$] (25)
                        -- (9) coordinate[c1,pos=0.8,label=90:${35}$] (35);

                \draw (7) -- (27) coordinate[c1,pos=0.5,label=90:${34}$] (34) coordinate[s1,label=90:${27}$] (27)
                        -- (23) coordinate[c1,pos=0.3,label=270:${26}$] (26);

                \draw  (15.5,17.5) coordinate[c1,label=15:${13}$] (13)
                        -- (29) coordinate[c1,pos=0.4,label=90:${31}$] (31);

                \draw (28) -- (10.5,15) coordinate[c1,pos=0.5,label=90:${30}$] (30) coordinate[c1,label=90:${15}$] (15)
                        -- (11) coordinate[s1,pos=0.6,label=90:${14}$] (14);

                \draw (15) -- (1) coordinate[s1,pos=0.2,label=15:${17}$] (17)
                        -- (13) coordinate[s1,pos=0.3,label=0:${16}$] (16);

                \draw (19) -- (16) coordinate[c1,pos=0.3,label=180:${18}$] (18)
                        -- (8) coordinate[c1,pos=0.7,label=180:${32}$] (32);

                \draw (6) -- (17) coordinate[c1,pos=0.7,label=90:${33}$] (33)
                        -- (21) coordinate[c1,pos=0.4,label=90:${20}$] (20);

                \draw (25) -- (1) -- (27);

                \coordinate (ContextLabel) at ([shift=({-2cm,-3mm})]1);

                \end{tikzpicture}
                }
\\
\\
(a)&&(b)
\end{tabular}
                \end{center}
                \caption{The TITS gadget hypergraph for $b$ given $v(a)=1$, as well as the TIFS gadget hypergraph for $3$ given $v(a)=1$,
which is partly reflection symmetric along the $\{a,1,2\}$ context to the TIFS gadget hypergraph in Figure~\ref{2023-viext-TIFS},
illustrates the orthogonality relations among vector labels of the elements of hyperedges~\cite{lovasz-79}
within a subset of quantum observables---also known as a faithful orthogonal representation~\cite{Portillo-2015} or coordinatization~\cite{Pavii2018},
as enumerated in~\cite[Table~I]{2018-minimalYIYS}. Red squares represent the value 1, and green circles represent the value 0.
(a) A single consistent value assignment is obtained by assuming $v(a)=1$ and $v(b)=1$ and applying the admissibility rules successively~\cite[Figure~(24.2.b)]{Svozil-2018-p}.
(b) An inconsistent value assignment is obtained by assuming $v(a)=1$ and $v(b)=0$ and applying the admissibility rules successively:
because the context $\{6,7,b\}$, shown dotted, contains three observables with the value 0,
no admissible value assignment $v$ exists with the constraint on the input and output ports $v(a)=1$ and $v(b)=0$.
Therefore, if $a$ has a preselected input state $v(a)=1$, then the value assignment $v(b)$.
For the output state, $b$ has to be 1 or undefined; that is, $b$ is an indefinite value.
}
                \label{2023-viext-TITS}
        \end{figure*}

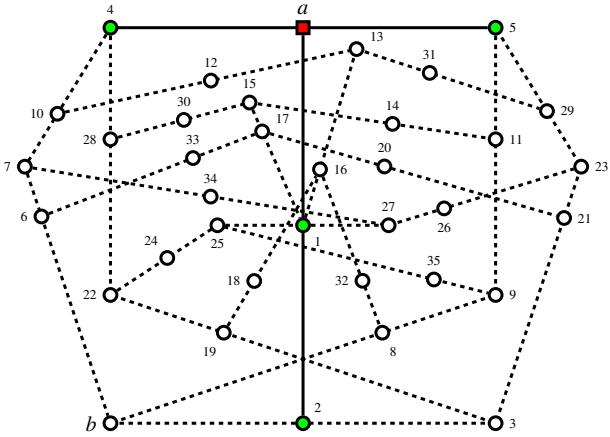
\begin{figure}
                \begin{center}
\resizebox{.46\textwidth}{!}{
                \begin{tikzpicture}  [scale=0.5]

                \tikzstyle{every path}=[line width=2pt]
                \tikzstyle{c1}=[draw,black,solid,fill=white,circle,inner sep={3},minimum size=6]
                \tikzstyle{cg1}=[draw,black,solid,fill=green,circle,inner sep={3},minimum size=6]
                \tikzstyle{s1}=[draw,black,solid,fill=red,rectangle,inner sep={4},minimum size=7]
                \tikzstyle{l1}=[draw=none,solid,circle,minimum size=4]


                \draw [dashed] (4,0) coordinate[c1,label=180:{\Large $b$}] (b)
                        -- (13,0) coordinate[cg1,label=45:$2$] (2)
                        -- (22,0) coordinate[c1,label=0:$3$] (3)
                        -- (26,12) coordinate[c1,pos=0.8,label=0:${21}$] (21) coordinate[c1,label=0:${23}$] (23)
                        -- (22,18.5) coordinate[c1,pos=0.4,label=0:${29}$] (29) coordinate[cg1,label=0:$5$] (5);
\draw (5)
                        -- (13,18.5) coordinate[s1,label=90:{\Large $a$}] (a)
                        -- (4,18.5) coordinate[cg1,label=90:$4$] (4);
\draw [dashed] (4)
                        -- (0,12) coordinate[c1,pos=0.6,label=180:${10}$] (10) coordinate[c1,label=180:$7$] (7)
                        -- (b) coordinate[c1,pos=0.2,label=180:$6$] (6);

                \draw (a) -- (2) coordinate[cg1,pos=0.5,label=315:$1$] (1);

                \draw [dashed] (5) -- (22,6) coordinate[c1,pos=0.4,label=0:${11}$] (11) coordinate[c1,label=0:$9$] (9)
                        -- (b) coordinate[c1,pos=0.3,label=280:$8$] (8);

                \draw [dashed] (4) -- (4,6) coordinate[c1,pos=0.4,label=180:${28}$] (28) coordinate[c1,label=180:${22}$] (22)
                        -- (3) coordinate[c1,pos=0.3,label=260:${19}$] (19);

                \coordinate (25) at ([xshift=-4cm]1);
                \coordinate (27) at ([xshift=4cm]1);

                \draw [dashed] (22) -- (25) coordinate[c1,pos=0.5,label=115:${24}$] (24) coordinate[c1,label=270:${25}$] (25)
                        -- (9) coordinate[c1,pos=0.8,label=90:${35}$] (35);

                \draw [dashed] (7) -- (27) coordinate[c1,pos=0.5,label=90:${34}$] (34) coordinate[c1,label=90:${27}$] (27)
                        -- (23) coordinate[c1,pos=0.3,label=270:${26}$] (26);

                \draw [dashed] (10) -- (15.5,17.5) coordinate[c1,pos=0.5,label=90:${12}$] (12) coordinate[c1,label=15:${13}$] (13)
                        -- (29) coordinate[c1,pos=0.4,label=90:${31}$] (31);

                \draw [dashed] (28) -- (10.5,15) coordinate[c1,pos=0.5,label=90:${30}$] (30) coordinate[c1,label=90:${15}$] (15)
                        -- (11) coordinate[c1,pos=0.6,label=90:${14}$] (14);

                \draw [dashed] (15) -- (1) coordinate[c1,pos=0.2,label=15:${17}$] (17)
                        -- (13) coordinate[c1,pos=0.3,label=0:${16}$] (16);

                \draw [dashed] (19) -- (16) coordinate[c1,pos=0.3,label=180:${18}$] (18)
                        -- (8) coordinate[c1,pos=0.7,label=180:${32}$] (32);

                \draw [dashed] (6) -- (17) coordinate[c1,pos=0.7,label=90:${33}$] (33)
                        -- (21) coordinate[c1,pos=0.4,label=90:${20}$] (20);

                \draw [dashed] (25) -- (1) -- (27);

                \coordinate (ContextLabel) at ([shift=({-2cm,-3mm})]1);

                \end{tikzpicture}
                }
                \end{center}
                \caption{Concatenated hypergraph from the hypergraphs depicting TIFS and TITS gadgets shown in Figures~\ref{2023-viext-TIFS} and~\ref{2023-viext-TITS},
respectively. Admissibility merely allows ``star-shaped'' value definite observables along the two contexts $\{a,1,2\}$ and $\{a,4,5\}$
if the system is prepared in state $a$.
}
                \label{2023-viext-TIFS&TITSconcatenated}
        \end{figure}

Conversely, the TITS gadget hypergraph in Figure~\ref{2023-viext-TITS}
illustrates the orthogonality relations among vector labels of the elements of hyperedges~\cite{lovasz-79}, as detailed in~\cite[Table~I]{2018-minimalYIYS}.
Using the admissibility rules~\cite[Fig.~(24.2.a)]{Svozil-2018-p} a single consistent value assignment,
as in Figure~(\ref{2023-viext-TITS}a) implies $v(a)=1$ and $v(b)=1$, in contrast with
the value assignment  when assuming $v(a)=1$ and $v(b)=0$.

As before, for any such configuration of quantum observables, there exists no classical admissible value assignment $v$
satisfying the constraint on the input and output ports $v(a)=1$ and $v(b)=0$, respectively.
Consequently, if $a$ has a preselected input state $v(a)=1$, then the value assignment $v(b)$ for the output state $b$ must be either 1 or undefined,
that is, value indefinite.

Therefore, the concatenation of the two hypergraphs depicting TIFS and TITS gadgets,
originally introduced by Abbott and the authors in~\cite{2018-minimalYIYS}, and
shown in Figures~\ref{2023-viext-TIFS} and~\ref{2023-viext-TITS} respectively, excludes both admissible value assignments of 0 and 1,
rendering $v(b)$ undefined and thus the observable $b$ value indefinite.
Indeed, as in Figure~\ref{2023-viext-TIFS&TITSconcatenated}
the penetration of admissible value assignments is rather limited: if the system is prepared in state $a$, then admissibility merely allows
``star-shaped'' value definite observables along the two contexts $\{a,1,2\}$ and $\{a,4,5\}$.
Note that all contexts
$\{b,2,3\}$,
$\{b,6,7\}$, and
$\{b,8,9\}$,
in which $b$ is an element, have at least one more element with indefinite value.
This is because
the set of observables $O=\{ a,b, 1,\ldots , 35\}$ is not unital~\cite{svozil-tkadlec}, that is, all eight admissible (or global) value assignments must assign the value 1 to the observable 1,
and thus the value 0 to $a$.
There does not exist any value assignment  $v(a)=1$~\cite[Table~24.1]{Svozil-2018-p}.
However, such value assignments with  $v(a)=1$ exist for the reduced set of observables
$O  \setminus \{29,31\}$
and
$O  \setminus \{10,12\}$
forming TIFS and TITS, respectively.

A very similar argument uses the same hypergraphs as in Figures~\ref{2023-viext-TIFS}
and~\ref{2023-viext-TITS} as TITS and TIFS gadgets for $3$ given $v(a)=1$, respectively.
Therefore, $v(3)$ is undefined, and the observable $3$ is value indefinite.

Finally, what are the effects of errors and system imperfections?
This question requires a technical long discussion, which will be the object of another study. Here, we argue only about the stability of the construction of our QRNGs due to variations in the indefinite observable value and measurement.

1. The stability of the choice of value indefinite observable comes from the Located Kochen-Specker Theorem~\cite{2012-incomput-proofsCJ,2015-AnalyticKS} stated at the beginning of this section:
The projection observable $P_\phi$ of {\it any state $\ket{\phi}$ such that {\rm $0<\abs{\bra{\psi}\ket{\phi}}<1$}}
is value
indefinite.

2. The stability of the measurement comes from the result proved at the beginning of this section, stating that any unitary operator is {\it universal}.

\section{Binary QRNG Based on Value Indefinite Observables}
\label{2023-viext-qrngdefs}

Subsequently, we present in detail an example of a configuration that illustrates a scenario where two observables within a context are value-indefinite, while the third observable is value-definite.

Here, value indefiniteness is contingent upon two factors:
(i) the state that is (pre-)selected and prepared, and
(ii) the specific set of observables arranged within a particular configuration of intertwined contexts.
To establish value indefiniteness within this configuration, the (pre-)selected state and the state characterized by value indefiniteness must be elements of the setup.
Therefore, any explicit assertion regarding the value indefiniteness of an observable should include a reference to the specific conditions upon which this claim relies.

\subsection{Quantum versus classical models}

A quantum realization of the construction  in Figures~\ref{2023-viext-TIFS},~\ref{2023-viext-TITS} and~\ref{2023-viext-TIFS&TITSconcatenated}
can be obtained from the faithful orthogonal representation of the elements of the hyperedges as vectors.
One such representation was given in~\cite[Table~I]{2018-minimalYIYS}.
It assigns the  (superscript $T$ indicates transposition)
$\vert a \rangle = \begin{pmatrix}1,0,0\end{pmatrix}^T$ to (the pure state) $a$,
also representable by the trace-class one orthogonal (that is, positive, self-adjoint) projection operator
whose matrix representation with respect to the Cartesian standard basis
is a diagonal matrix
$E_a=\vert a \rangle \langle a \vert = \text{diag}\begin{pmatrix} 1,0,0 \end{pmatrix}^T$
and
$\vert b \rangle = \begin{pmatrix}\frac{1}{\sqrt{2}},\frac{1}{2},\frac{1}{2}\end{pmatrix}^T$
as well as
$\vert 3 \rangle = \begin{pmatrix}\frac{1}{\sqrt{2}},-\frac{1}{2},-\frac{1}{2}\end{pmatrix}^T$
to the observables $b$ and $3$, respectively.
Therefore, if the system is preselected (or prepared) in state $\vert a \rangle$, the output of the
measurement of
\begin{equation*}
E_b=\vert b \rangle \langle b \vert =  \frac{1}{2}
\begin{pmatrix}
1&\frac{1}{\sqrt{2}}&\frac{1}{\sqrt{2}} \\
\frac{1}{\sqrt{2}}&\frac{1}{2}&\frac{1}{2} \\
\frac{1}{\sqrt{2}}&\frac{1}{2}&\frac{1}{2}
\end{pmatrix}
\end{equation*}
along  $\vert b \rangle$
 is obtained with the probability
\begin{equation*}
\text{Tr}\big(E_a \cdot E_b \big) = \vert \langle b \vert a \rangle \vert^2 = \frac12.
\end{equation*}
Likewise, the output of the
measurement of
\begin{equation*}
E_3=\vert 3 \rangle \langle 3 \vert =  \frac{1}{2}
\begin{pmatrix}
1&-\frac{1}{\sqrt{2}}&-\frac{1}{\sqrt{2}} \\
-\frac{1}{\sqrt{2}}&\frac{1}{2}&\frac{1}{2} \\
-\frac{1}{\sqrt{2}}&\frac{1}{2}&\frac{1}{2}
\end{pmatrix}
\end{equation*}
along  $\vert b \rangle$
 is obtained with the probability

\begin{equation*}
\text{Tr}\big(E_a \cdot E_3 \big) = \vert \langle 3 \vert a \rangle \vert^2 = \frac12.
\end{equation*}

As $\vert 2 \rangle$ is orthogonal to $\vert a \rangle$,
$\text{Tr}\big(E_a \cdot E_2 \big) = \vert \langle 2 \vert a \rangle \vert^2 = 0$,
and the observable $2$ is defined.
Hence, when the observable $a$ is preselected in the state $\vert a \rangle$, both observables $b$ and $3$ become value-indefinite (relative to admissibility),
while observable $2$ has value  $v(2) = 0$.
A quantum calculation confirmes what is posited in the (Located) Kochen-Specker Theorem,
that both $b$ and $3$ occur with a probability of $\frac12$.

To emphasize the three-dimensionality of the configuration, even if only two observables have nonzero probabilities,  the sum of frequencies of the remaining quantum observables $2$ and $3$
 in the complement  $\{2,3\}$ of the context $\{b,2,3\}$ containing $b$ is $1/2$.
More explicitly, expressed in terms of orthogonal projection operators, the observable corresponding to $\{2,3\}$
is given by a  matrix  corresponding to the orthogonal projection operator $E_{2,3}$:
\begin{equation}
\begin{split}
E_{2,3}=E_2 + E_3 = \vert 2 \rangle \langle 2 \vert + \vert 3 \rangle \langle 3 \vert \\
\quad = \frac12
\begin{pmatrix}
1 & -\frac1{\sqrt{2}}& -\frac1{\sqrt{2}}\\
-\frac1{\sqrt{2}}& \frac32& -\frac12\\
-\frac1{\sqrt{2}}& -\frac12& \frac32
\end{pmatrix}
.
\end{split}
\label{2023-viext-sumotopo}
\end{equation}
The vectors in $E_{2,3}\in \C^3$ are orthogonal to vectors in  $E_{b}\in \C^3$.
Together, $E_b+E_{2,3} =\vert b \rangle \langle b \vert +\vert 2 \rangle \langle 2 \vert + \vert 3 \rangle \langle 3 \vert =I_3$ yield
the identity $I_3=\text{diag}\begin{pmatrix}1,1,1\end{pmatrix}$.

Classically, there is no realization of the set of observables $O=\{ a, b, 1,\ldots , 35\}$ in Figure~\ref{2023-viext-TIFS&TITSconcatenated} because
some elements of $O$ are assigned the value $0$ for all two-valued states~\cite[Table~24.1]{Svozil-2018-p}, hence
not separable~\cite[Theorem~0]{kochen1}.  This result holds for total value assignments---a stronger assumption than admissibility.
Indeed, in this case  the ``central'' point 1 must be classically assigned the value $v(1)=1$,
and, therefore, all remaining eight elements
$\{a, 2, 13, 15, 16, 17, 25, 27\}$
in the four contexts
$\{a,1,2\}$, $\{1,13,16\}$, $\{1,15,17\}$, and $\{1,25,27\}$ containing 1 to be zero.

Finally, using the Eigenstate principle
and Theorem 5.6 in~\cite{RSPA23}, we deduce that the QRNG described above generates maximally unpredictable binary random digits.
\subsection{Beam splitter realizations}

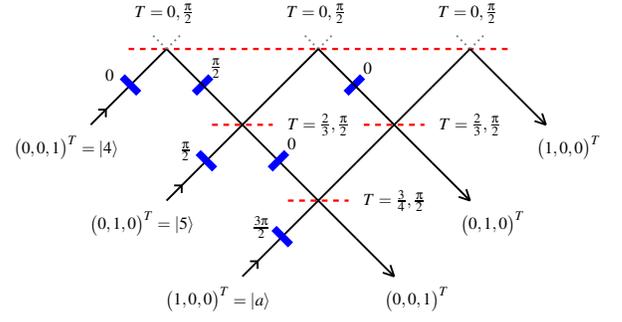
\begin{figure}[ht]
\begin{center}
\resizebox{.45\textwidth}{!}{
\begin{tikzpicture} [scale=2, rotate=45]
\tikzstyle{every path}=[line width=1pt]
\draw [red,dashed,very thick] ({1-0.25},{2+0.25}) -- ({3+0.25},{0-0.25});
\node at (1.22,2.25)  {$T=0, \frac{\pi}{2}$};
\node at (2.22,1.25)  {$T=0, \frac{\pi}{2}$};
\node at (3.22,0.25)  {$T=0, \frac{\pi}{2}$};
\node at (1.5,0.5)  {$T=\frac{2}{3}, \frac{\pi}{2}$};
\draw [red,dashed,very thick] (0.8,1.2) -- (1.2,1 - 0.2 );
\draw [fill,blue] (0.9,1.5) rectangle ++(0.2,0.05);
\node at (1.2,1.55) {$\frac{\pi}{2}$};
\draw [black] (0.5,1) -- (1.5,1);
\draw [black] (1,0.5) -- (1,1.5);
\node at (1.5,-0.5)  {$T=\frac{3}{4}, \frac{\pi}{2}$};
\draw [red,dashed,very thick] (0.8,0.2) -- (1.2,-0.2 );
\draw [fill,blue] (0.9,0.5) rectangle ++(0.2,0.05);
\node at (1.2,0.55) {$0$};
\draw [black] (0.5,0) -- (1.5,0);
\draw [black] (1,-0.5) -- (1,0.5);
\node at (2.5,-0.5)  {$T=\frac{2}{3}, \frac{\pi}{2}$};
\draw [red,dashed,very thick] (1.8,0.2) -- (2.2,-0.2 );
\draw [fill,blue] (1.9,0.5) rectangle ++(0.2,0.05);
\node at (2.2,0.55) {$0$};
\draw [black] (1.5,0) -- (2.5,0);
\draw [black] (2,-0.5) -- (2,0.5);
\draw [black] (0,2) -- (0.5,2);
\node at (-0.2,1.9) {$\begin{pmatrix} 0, 0, 1 \end{pmatrix}^T = \vert 4 \rangle\qquad$};
\draw [black] (3,-0.5) -- (3,-1);
\node at (3,-1.3) {$\begin{pmatrix}1,0,0\end{pmatrix}^T$};
\draw [black] (0.5,2) -- (1,2);
\draw [black] (1,2) -- (1,1.5);
\draw [black] (0,1) -- (0.5,1);
\node at (-0.2,0.9) {$\begin{pmatrix} 0, 1, 0\end{pmatrix}^T = \vert 5 \rangle\qquad$};
\draw [black] (2,-0.5) -- (2,-1);

(* arrows*)
\draw [black] (3,-1) -- (2.95,-0.9);
\draw [black] (3,-1) -- (3.05,-0.9);
\draw [black] (2,-1) -- (1.95,-0.9);
\draw [black] (2,-1) -- (2.05,-0.9);
\draw [black] (1,-1) -- (0.95,-0.9);
\draw [black] (1,-1) -- (1.05,-0.9);
\draw [black] (0.2,1) -- (0.15,1.05);
\draw [black] (0.2,1) -- (0.15,0.95);
\draw [black] (0.2,0) -- (0.15,0.05);
\draw [black] (0.2,0) -- (0.15,-0.05);
\draw [black] (0.2,2) -- (0.15,2.05);
\draw [black] (0.2,2) -- (0.15,1.95);

(* symmetric beamsplitter over-extension *)

\draw [gray,dotted] (1,2) -- (1.2,2);
\draw [gray,dotted] (1,2) -- (1,2.2);

\draw [gray,dotted] (2,1) -- (2.2,1);
\draw [gray,dotted] (2,1) -- (2,1.2);

\draw [gray,dotted] (3,0) -- (3.2,0);
\draw [gray,dotted] (3,0) -- (3,0.2);

\node at (2,-1.3) {$\begin{pmatrix}0,1,0\end{pmatrix}^T$};
\draw [black] (1.5,1) -- (2,1);
\draw [black] (2,1) -- (2,0.5);
\draw [black] (0,0) -- (0.5,0);
\node at (-0.2,-0.1) {$\begin{pmatrix} 1, 0, 0 \end{pmatrix}^T = \vert a \rangle\qquad$};
\draw [black] (1,-0.5) -- (1,-1);
\node at (1,-1.3) {$\begin{pmatrix}0,0,1\end{pmatrix}^T$};
\draw [black] (2.5,0) -- (3,0);
\draw [black] (3,0) -- (3,-0.5);
\draw [fill,blue] (0.5,-0.1) rectangle ++(0.05,0.2);
\node at (0.45,0.2) {$\frac{3\pi}{2}$};
\draw [fill,blue] (0.5,0.9) rectangle ++(0.05,0.2);
\node at (0.45,1.2) {$\frac{\pi}{2}$};
\draw [fill,blue] (0.5,1.9) rectangle ++(0.05,0.2);
\node at (0.45,2.2) {$0$};
\end{tikzpicture}
}
\end{center}
\caption{A triangular array of quantum mechanical beam splitters is a realization of the input or preparation context $\{a,4,5\}$ and the output or measurement context $\{b,2,3\}$.
\label{2023-viext-bsr}}
\end{figure}

Figure~\ref{2023-viext-bsr} presents a triangular array of quantum  beam splitters
 which physically transforms the preparation context  $\{a,4,5\}$
into the measurement context $\{b,2,3\}$.

The vector coordinatization ~\cite[Table~I]{2018-minimalYIYS}
$\vert a \rangle = \begin{pmatrix} 1, 0, 0\end{pmatrix}^T$,
$\vert b \rangle = \begin{pmatrix} \frac1{\sqrt{2}}, \frac12, \frac12 \end{pmatrix}^T$,
$\vert 2 \rangle = \begin{pmatrix} 0, \frac1{\sqrt{2}}, -\frac1{\sqrt{2}} \end{pmatrix}^T$,
$\vert 3 \rangle = \begin{pmatrix} \frac1{\sqrt{2}}, -\frac12, -\frac12 \end{pmatrix}$,
$\vert 4 \rangle = \begin{pmatrix} 0, 0, 1\end{pmatrix}^T$, and
$\vert 5 \rangle = \begin{pmatrix} 0, 1, 0\end{pmatrix}^T$
computes the unitary transformation matrix~\cite{Schwinger.60,Joglekar-I}
that transforms the input state $\vert a\rangle$ into the output state $\vert b\rangle$,
 the input state $\vert 4\rangle$ into the output state $\vert 2\rangle$,
and  the input state $\vert 5\rangle$ into the output state $\vert 3\rangle$:

\begin{equation}
\begin{split}
U =
\vert b \rangle \langle a \vert
+
\vert 2 \rangle \langle 4 \vert
+
\vert 3 \rangle \langle 5 \vert  \\
\quad =  \frac12
\begin{pmatrix}
\sqrt{2} &  \sqrt{2}& 0\\
1& -1&  \sqrt{2} \\
1& -1& - \sqrt{2}
\end{pmatrix}
.
\end{split}
\label{2023-viext-uo}
\end{equation}

This unitary matrix realizes  a beam splitter~\cite{reck-94,rzbb,
de_Guise_2018}
using the parametrization of the unitary group~\cite{murnaghan}.
Besides phase shifters operating in one-dimensional subspaces (in this particular case, all zero but one),
these concatenations of optical elements
contain beam splitters operating in two-dimensional subspaces.
These beam splitters have a parametrization unitary matrix
\begin{equation*}
B( \omega , \varphi )  =
\begin{pmatrix}
 \sin  \omega  & \cos  \omega
\\
e^{-i \varphi } \cos  \omega  &-e^{-i \varphi } \sin  \omega
\end{pmatrix}
\end{equation*}
depending on two parameters:
 $\omega$ is the transmissivity $T=\sin^2\omega$ and reflectivity $R=1-T=\cos^2\omega$,
and $\varphi$ is the phase change at reflection.

The output wave function, given the input wave function,
is the coherent superposition of the contributions of all the possible forward passes from the input port(s) toward the output port(s).
Thereby, the transmissibility and reflectivity contribute by the square roots $\sqrt{T}=\sin \omega$
and reflectivity $\sqrt{R}=\cos \omega$ of $T$ and $R$~\cite{green-horn-zei}.
The sum of the phase shifts between reflected and transmitted waves
excited by a wave incident from the side of the beam splitter, and the corresponding phase shift
for a wave incident from the opposing side, contribute with $\pi$~\cite{zeilinger:882}.
For a symmetric lossless dielectric plate~\cite{Lai_1985},
the reflected and transmitted parts are $\pi/2$ out of phase~\cite{Degiorgio_1980,green-horn-zei}.

The relations~(\ref{2023-viext-owf}) present a computation of the effects on the input ports of the beam splitter
in Figure~\ref{2023-viext-bsr} by successive applications of phase shifts and beam mixings.
\begin{widetext}
\begin{equation}
\begin{split}
\vert a \rangle
\longrightarrow
e^{i\frac{3\pi}{2}} \left\{
 e^{i\frac{\pi}{2}} \sqrt{\frac14}  \begin{pmatrix}0\\0\\1\end{pmatrix} +   \sqrt{\frac34}  \left[  e^{i\frac{\pi}{2}}  \sqrt{\frac13}  \begin{pmatrix}0\\1\\0\end{pmatrix} +    \sqrt{\frac23}    e^{i\frac{\pi}{2}} \begin{pmatrix}1\\0\\0\end{pmatrix} \right]
\right\} =  \vert b \rangle
,
\\
\vert 5 \rangle
\longrightarrow
 e^{i\frac{\pi}{2}} \left(
 e^{i\frac{\pi}{2}}  \sqrt{\frac13} \left\{
 \sqrt{\frac34}   \begin{pmatrix}0\\0\\1\end{pmatrix} +  e^{i\frac{\pi}{2}}  \sqrt{\frac14} \left[
                                            e^{i\frac{\pi}{2}}  \sqrt{\frac13}  \begin{pmatrix}0\\1\\0\end{pmatrix} +  e^{i\frac{\pi}{2}}  \sqrt{\frac23}  \begin{pmatrix}1\\0\\0\end{pmatrix}
                                           \right]
                      \right\}  \right.
\\ \qquad \qquad
\left.
+  \sqrt{\frac23}   e^{i\frac{\pi}{2}}  \left[   \sqrt{\frac23}  \begin{pmatrix}0\\1\\0\end{pmatrix} +  e^{i\frac{\pi}{2}}  \sqrt{\frac13}   e^{i\frac{\pi}{2}} \begin{pmatrix}1\\0\\0\end{pmatrix}
                      \right]
\right)
=  \vert 3 \rangle
,
\\
\vert 4 \rangle
\longrightarrow
 e^{i\frac{\pi}{2}}  e^{i\frac{\pi}{2}} \left(
             \sqrt{\frac23} \left\{
                      \sqrt{\frac34}  \begin{pmatrix}0\\0\\1\end{pmatrix} +  e^{i\frac{\pi}{2}}  \sqrt{\frac14}   \left[
                                                                  e^{i\frac{\pi}{2}}  \sqrt{\frac13}  \begin{pmatrix}0\\1\\0\end{pmatrix} +   \sqrt{\frac23}   e^{i\frac{\pi}{2}} \begin{pmatrix}1\\0\\0\end{pmatrix}
                                                                \right]
                     \right\}
\right.
\\ \qquad \qquad
\left.
            +  e^{i\frac{\pi}{2}}  \sqrt{\frac13}   e^{i\frac{\pi}{2}}                 \left[
                                                                  \sqrt{\frac23}  \begin{pmatrix}0\\1\\0\end{pmatrix} +  e^{i\frac{\pi}{2}}  \sqrt{\frac13}   e^{i\frac{\pi}{2}} \begin{pmatrix}1\\0\\0\end{pmatrix}
                                                                \right]
             \right)
=  \vert 2 \rangle
.
\end{split}
\label{2023-viext-owf}
\end{equation}
\end{widetext}

\section{Beam splitter as an analogy of Ariadne's tread}

How come can we quantum mechanically `spread' a  qutrit state of input  into a coherent superposition of all qutrit states,
and finally end up with a binary sequence---very much like two Hadamard unitary transformations first `spread' a qubit,
and then (up to a constant scalar factor) `fold it back' into its original state?
This is where the allegory of Ariadne's thread comes up in the configuration of a beam splitter.
Consider a general quantum beam splitter with $m>0$ nonzero inputs and $n>0$ nonzero output ports.
As long as the sum of probabilities of preparation and detection on both the respective input and the output ports adds up to one,
a quantum realization is feasible~\cite{reck-94,rzbb,
de_Guise_2018}. Indeed, all that is necessary
is that the input and the output state are tailored according to the probability amplitudes (phases do not count).

Considering this scenario, one may question:
What happens to quantum unitarity, especially if $m \neq n$?
For instance, with such a beam splitter, we could `merge' two input ports into one output port ($n=m+1=2$).
 Alternatively, one could `split' a single input port into (a coherent superposition, resulting in) two output ports ($m=n+1=2$).
For example, the associated unitary three-dimensional matrix entries could be
\begin{equation}
U_{\text{2-to-1}}=
\begin{pmatrix}
0&\frac{1}{\sqrt{2}}&\frac{1}{\sqrt{2}} \\
\cdot&\cdot&\cdot \\
\cdot&\cdot&\cdot
\end{pmatrix}
,\phantom{x}
U_{\text{1-to-2}}=
\begin{pmatrix}
0&\cdot&\cdot \\
\frac{1}{\sqrt{2}}&\cdot&\cdot \\
\frac{1}{\sqrt{2}}&\cdot&\cdot
\end{pmatrix},
\end{equation}
where, for $U_{\text{2-to-1}}$ (or $U_{\text{1-to-2}}$)
the remaining rows (or columns) could fill up with unit vectors forming the orthonormal basis of a two-dimensional subspace orthogonal to
$
\begin{pmatrix}
0&\frac{1}{\sqrt{2}}&\frac{1}{\sqrt{2}}
\end{pmatrix}
$
(or its Hermitian conjugate).

Indeed, to obtain a binary sequence, one could `post-process' the beam splitter arrangement  in
Figure~\ref{2023-viext-bsr} by a beam splitter corresponding to the following real-valued unitary matrix:
\begin{equation}
U'_{\text{2-to-1}}=
\frac{1}{\sqrt{2}}
\begin{pmatrix}
\sqrt{2}&0&0\\
0&1&1 \\
0&1&-1
\end{pmatrix}
.
\label{2023-viext-ub1to}
\end{equation}

When the input state is $\vert a \rangle$, the resulting output state is  $U'_{\text{2-to-1}}U\vert a \rangle$, with $U$ and $U'_{\text{2-to-1}}$ defined in~(\ref{2023-viext-u}) and~(\ref{2023-viext-ub1to}), respectively.

More explicitly,
\begin{equation*}
\begin{split}
\frac{1}{\sqrt{2}}
\begin{pmatrix}
\sqrt{2}&0&0\\
0&1&1 \\
0&1&-1
\end{pmatrix}
 \frac12
\begin{pmatrix}
\sqrt{2} &  \sqrt{2}& 0\\
1& -1&  \sqrt{2} \\
1& -1& - \sqrt{2}
\end{pmatrix}
\begin{pmatrix}
1\\
0 \\
0
\end{pmatrix}
=
 \frac{1}{\sqrt{2}}
\begin{pmatrix}
1\\
1 \\
0
\end{pmatrix}
.
\end{split}
\end{equation*}
 A particle in state $\vert a \rangle$ will end up in either the first or second port with probability $\frac{1}{2}$ and be registered in the third port with probability 0.

Two questions arise:
(i)
The unitary quantum evolution---of the von Neumann type `Vorgang' 2~\cite{v-neumann-49,v-neumann-55}, referred to as `process 2' by Everett~\cite{everett}---that
needs to be one-to-one, and it appears to be compromised.
(ii)
Can this problem be discussed in terms of value indefiniteness and partial value assignments?

The first question can be quickly addressed:
The beam splitter examples discussed here show that concentration on a partial array of input and output ports cannot represent the whole picture.
The full specification of a beam splitter in $n$ dimensions has the same number $n$ of input and output ports.
The quantum evolution is incomplete if some input and output contexts are not considered.
Because any unitary transformation can be represented by a bijective map of the vectors
of one orthonormal basis---the input context---into the vectors of another orthonormal basis~\cite{Schwinger.60,Joglekar-I}---the output context.
Suppose we also allow incomplete mappings of vectors from one context into some vectors of another context. This could not exclude mappings that are not one-to-one. Therefore,
only the totality of those one-to-one vector transformations relating to two orthonormal bases forms a forward- and backward-reversible transformation.

The context-to-context unitary mapping can be viewed as a sort of `rescrambling'
of information contained in the channels or ports of the beam splitter~\cite{schrodinger-gwsidqm2,zeil-99}.
Thereby, the `latent' and `omitted' ports act as Ariadne's thread that must be considered for reversibility.
The situation resembles a zero-sum game encountered in
entanglement swapping~\cite{BBCJPW,peres-DelayedChoiceEntanglementSwapping}.

Although the results in this article have been proved in  $\C^3$, they can easily be generalized to $\C^n$ with $ n>3$.
Therefore, by `merging'
or `folding'
two or more observables of the context, represented by the orthogonal projection operators $E_2,\ldots , E_n$,
we never leave the $n$-dimensional Hilbert space $\C^n$,
because $E_{2,\ldots,n}\C^n$ is the $(n-1)$-dimensional Hilbert space spanned by the vectors
$\vert e_i\rangle$ that form
$E_i=\vert e_i\rangle \langle e_i \vert$, with $i=2,\ldots, n$.
The vectors in $E_{2,\ldots,n}\C^n$ are orthogonal to the one-dimensional subspace $E_{1}\C^n$ spanned by $\vert e_1\rangle$,
and the vectors $\vert e_1\rangle,\ldots ,\vert e_1\rangle$ form an orthonormal basis.

Regarding the second question, we may say that value indefiniteness `prevails' over value definiteness: whenever a value indefinite observable is involved, the `merged' observables `inherit' value indefiniteness.

\section{Conclusions}

We have proved that
for every probability distribution $(p_1,p_2,p_3)$ ($\sum_{i}p_i=1$ and $0\le  p_i < 1$), one can construct a value indefinite quantum state which, by {\it every unitary  measurement}, produces the
outcomes with probabilities $p_i$.

Based on this result, the quantization of an algorithmic pre-processing binary method~\cite{CALUDE202131} and the quantum `merging' technique, we have constructed quantum random generators based on measuring a three-dimensional value indefinite observable producing binary quantum random outputs with the same randomness qualities as the ternary ones; their outputs are maximally unpredictable~\cite{2014-nobit}. The results can easily be generalized from $\C^3$ to $\C^n$ with $ n>3$.

\begin{acknowledgments}
We thank J.~M. Ag\"{u}ero Trejo for comments, which improved the presentation, and M.~Reck for the Mathematica code producing the generalized beam-splitter setup for an arbitrary unitary transformation.
The research of K.~Svozil was funded in whole or in part by the Austrian Science Fund (FWF), Project No. I 4579-N.
\end{acknowledgments}

\bibliography{svozil}

\begin{thebibliography}{50}
\bibitem[{\citenamefont{Fedorov}(2022)}]{Arkady_Fedorov-pc}
\bibinfo{author}{\bibfnamefont{A.}~\bibnamefont{Fedorov}},
  \emph{\bibinfo{title}{Binary beam splitters may not be protected by value
  indefiniteness as tertiary ones}} (\bibinfo{year}{2022}),
  \bibinfo{note}{private communications}.
\end{thebibliography}

\end{document}